\documentclass[]{spie}  

\usepackage{amsmath,amsfonts,amssymb}
\usepackage{graphicx}
\usepackage[colorlinks=true, allcolors=blue]{hyperref}
\usepackage{multirow}
\usepackage{subfigure}

\title{Reducing the background in X-ray imaging detectors via machine learning}

\author[a]{D.~R.~Wilkins}
\author[a]{S.~W.~Allen}
\author[b]{E.~D.~Miller}
\author[b]{M.~Bautz}
\author[a]{T.~Chattopadhyay}
\author[b]{R.~Foster}
\author[b]{C.~E.~Grant}
\author[a]{S.~Herrmann}
\author[c]{R.~Kraft}
\author[a]{R.~G.~Morris}
\author[c]{P.~Nulsen}
\author[c]{G.~Schellenberger}
\affil[a]{Kavli Institute for Particle Astrophysics and Cosmology, Stanford University, 452 Lomita Mall, Stanford, CA 94305, USA}
\affil[b]{MIT Kavli Institute for Astrophysics and Space Research, 77 Massachusetts Avenue, Cambridge, MA 02139, USA}
\affil[c]{Harvard–Smithsonian Center for Astrophysics, 60 Garden Street, Cambridge, MA 02138, USA}

\authorinfo{Correspondence e-mail: dan.wilkins@stanford.edu}

\begin{document} 
\maketitle

\begin{abstract}
The sensitivity of astronomical X-ray detectors is limited by the instrumental background. The background is especially important when observing low surface brightness sources that are critical for many of the science cases targeted by future X-ray observatories, including Athena and future US-led flagship or probe-class X-ray missions. Above 2\,keV, the background is dominated by signals induced by cosmic rays interacting with the spacecraft and detector. We develop novel machine learning algorithms to identify events in next-generation X-ray imaging detectors and to predict the probability that an event is induced by a cosmic ray vs. an astrophysical X-ray photon, enabling enhanced filtering of the cosmic ray-induced background. We find that by learning the typical correlations between the secondary events that arise from a single primary, machine learning algorithms are able to successfully identify cosmic ray-induced background events that are missed by traditional filtering methods employed on current-generation X-ray missions, reducing the unrejected background by as much as 30 per cent.
\end{abstract}


\section{Introduction}
\label{sec:intro}  
Highly sensitive X-ray imaging detectors are central to many of the scientific objectives of the next generation X-ray observatories, including the flagship European X-ray observatory, \textit{Athena}\cite{athena}, the NASA probe class mission concept \textit{AXIS}\cite{axis}, or a future US flagship X-ray mission\cite{lynx} as recommended for development by the \textit{Astro 2020} Decadal Survey\cite{decadal}. X-ray imaging detectors are most commonly based on CCD or CMOS technology, or DEPFET technology in the case of the \textit{Athena Wide Field Imager} (WFI)\cite{wfi}. Photons absorbed within pixels of the silicon device deposit energy, leading to the accumulation of charge within those pixels. After some integration time, a frame is read out of the detector; essentially an image of the charge accumulated per pixel during each integration time, from which the X-ray events that were detected can be reconstructed.

The sensitivity of astronomical X-ray detectors above 2\,keV is limited by the instrumental background: the accumulation of charge and the induction of signals within pixels of the detector by events other than `genuine' astrophysical X-ray photons that have travelled through the telescope optics to the detector. In this energy range, the instrumental background is dominated by cosmic rays (often referred to as the \textit{non-X-ray background}, NXB, or \textit{particle background}). Charged particles can interact directly with the detector, depositing energy and inducing signals in a series of pixels as they pass through. Alternatively, those same charged particles can interact with other parts of the spacecraft, where they may produce a shower of secondary particles, including further protons, electrons and positrons, and X-ray photons produced by fluorescent line emission in the materials making up the spacecraft. These secondaries can then go on to reach the detector, inducing their own signals, many of which can be confused with astrophysical X-rays, creating a background signal\cite{wfi_bkg}. The instrumental background is particularly critical when observing low surface brightness sources, such as the outskirts of galaxy clusters, which yield vast amounts of information about the formation of structure in the Universe, or when performing surveys to find active galactic nuclei (AGN) at high redshift, to learn about the seeds that formed supermassive black holes, their growth, and the role they played in the formation of galaxies\cite{athena_science}.

X-ray (and other) events must be reconstructed from the raw frames, containing the energy or charge deposited per pixel. Typically, this is done using an event grading scheme first employed on the ASCA satellite (henceforth referred to as the `ASCA grading' scheme). An event is defined as a $3\times 3$ pixel island, surrounding the pixel with the local maximum signal that must exceed the defined `event threshold.' The event is then assigned a grade, based upon the number of pixels within the island that are above a lower `split threshold' as well as the spatial arrangement of those pixels. The event energy is obtained by summing the energy recorded in all of the pixels that are above the threshold.

When an X-ray photon is absorbed within the silicon, a cloud of electrons is created that diffuses outwards before being collected at the electrode(s). Depending on the size and depth of the pixel, and whereabouts the photon is absorbed relative to the edges of the pixel, not all of that charge may be contained within a single pixel, and an X-ray event can be split between multiple, neighboring pixels\cite{miller_charge_diff}. When the pixel size is relatively large (for example in the \textit{XMM-Newton} EPIC pn CCD, or the \textit{Athena} WFI), the vast majority of X-rays will be recorded as single or double pixel events. When the pixels are smaller, such as in the case of the \textit{Chandra} ACIS CCDs, quadruple-pixel events are also common.

On the other hand, a proton with sufficiently high energy that it is considered a \textit{minimum ionizing particle} (MIP) will not be absorbed in a single pixel, but will rather traverse the detector, depositing energy in a number of pixels along its track. Charged particle events can readily be filtered from the event list on the basis of either an `invalid pattern' (\textit{i.e.} an extended track, rather than single, double, or sometimes quadruple pixel events) or `invalid energy' (above the MIP threshold) where the energy deposited in a single pixel exceeds the maximum that could be carried by a single photon reaching the detector via the mirrors.

After such filtering, however, there remains significant background signal of `valid events' that are induced by cosmic rays, including secondary electrons and fluorescent X-ray photons, both of which deposit relatively small amounts of energy in single pixels. If these secondary events can be associated with a primary proton track, this component of the background can be reduced. A recently-developed procedure, known as self-anticoincidence\cite{miller_jatis}, achieves this by masking areas of the detector around extended proton tracks. However, this comes at the cost of reducing the effective area of the detector, and can only remove secondary, valid events, when they appear close to a primary proton track in the same frame, which is not always present.

We explore whether machine learning algorithms can be trained to improve the filtering of the cosmic ray induced background in CCD, CMOS and DEPFET X-ray imaging detectors. Machine learning algorithms can be trained to interpret an image frame holistically, rather than considering $3\times 3$ pixel islands, to optimally segment the frame into the pixels that arise from single photon or charged particle events, and can learn to associate multiple secondary particles that appear in a single frame from a single cosmic ray primary, potentially reducing the unrejected component of the instrumental background. Machine learning algorithms can be set up to provide a probabilistic classification of each event, yielding the probability that any given event is induced by a cosmic ray, enabling flexible filtering criteria to be defined during the scientific analysis of the data.

The development of these algorithms is based upon \textsc{geant4} simulations of the interactions between cosmic ray protons and the spacecraft and detector, described in \S\ref{sec:sims}. We discuss two components of a machine learning algorithm, the first to classify frame images and identify regions of interest (\S\ref{sec:frame_class}), and the second to classify individual pixels within an image (\S\ref{sec:pixel_class}). We then discuss the performance of a prototype algorithm that combines these techniques in \S\ref{sec:hybrid}.

\section{Simulating cosmic ray background events}
\label{sec:sims}
We develop algorithms to detect and filter the component of the instrumental background that is induced by cosmic rays, based upon simulations of the interactions between cosmic ray particles and the \textit{Athena Wide Field Imager} \cite{grant+2020,miller_jatis}. Simulations were conducted in \textsc{geant4}\cite{geant4} as part of the AREMBES project\cite{wfi_bkg}, which aims to study and mitigate the background on the \textit{Athena} X-ray observatory.

The simulations trace cosmic ray protons through a mass model of the \textit{Athena} spacecraft and the WFI instrument. The WFI detector itself is defined within the mass model as a sheet of silicon, upon which a co-ordinate system of pixels is defined. The energy of the protons is sufficiently high that they are considered to be minimum ionising particles (MIPs), meaning that they are not stopped in the silicon `detector' but rather deposit energy as they pass through. The total energy deposited within each pixel of the silicon is recorded.

\begin{figure}
    \centering
    \small
    \subfigure[] {
    \includegraphics[height=7cm]{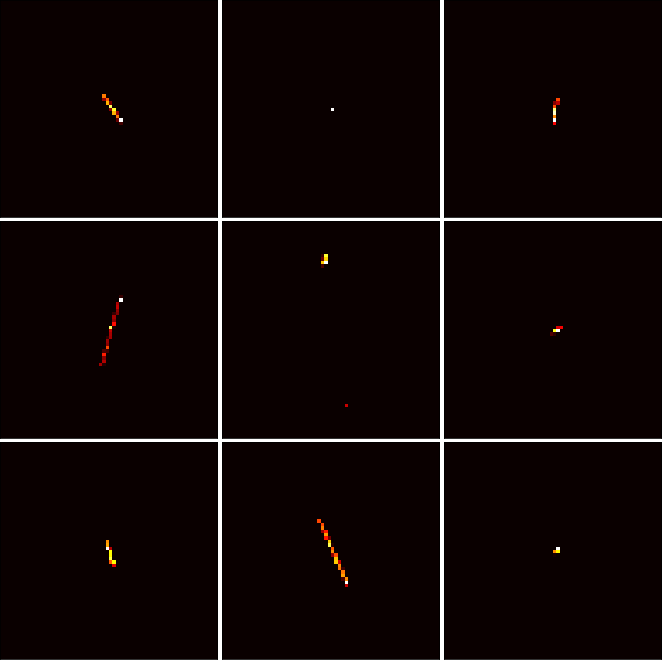}
    \label{fig:sims:gcr}
    }
    \subfigure[] {
    \includegraphics[height=7cm]{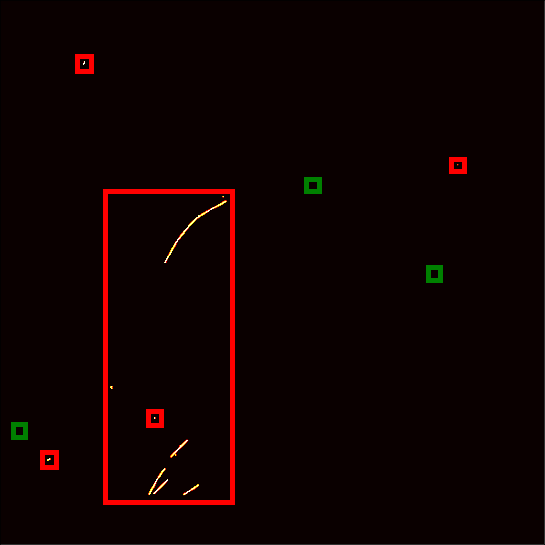}
    \label{fig:sims:frame}
    }
    \caption{\subref{fig:sims:gcr} \textsc{geant4} simulations of interactions between cosmic ray particles and the \textit{Athena} WFI detector. The simulation traces cosmic ray protons and their secondaries as they pass through and interact with the spacecraft, and the energy deposited by these particles in each of the pixels of the detector is recorded. \subref{fig:sims:frame} Simulated frame read out from a single $512\times 512$ pixel quadrant of the WFI detector after a single integration time. Event rates are enhanced for this simulation. Red boxes indicate cosmic ray events, each encompassing the primary protons and all of their secondaries. Green boxes indicate astrophysical X-ray events (photons that reach the detector via the telescope optics).}
    \label{fig:sims}
\end{figure}

Alternatively, the primary protons may interact with other parts of the spacecraft within the mass model. During each of these interactions, they may produce secondary particles, including further protons, electrons and positrons, and X-ray photons that are produced by fluorescence within the  material. Any of these secondaries may subsequently go on to reach the detector, themselves depositing energy into the pixels. For the purposes of this work, we consider a cosmic ray `event' to be the sum of the energy deposited in all of the pixels by both the primary proton and all of the secondary particles associated with it. A library of \textsc{geant4} simulations has been created, containing the energy deposited per pixel as a result of individual primary cosmic ray protons (Figure~\ref{fig:sims}a).

We assume that the detector is read out frame by frame, and that following some integration time (nominally 5\,ms per frame for the \textit{Athena} WFI), the signal recorded in each pixel is directly proportional to the energy deposited in that pixel during that time. We simulate frames that would be read out from each $512 \times 512$ pixel quadrant of the WFI detector. 

In addition to the cosmic ray signals, the simulated frames include astrophysical X-ray signals, that is the `genuine' X-ray photons that reach the detector via the telescope optics from celestial sources. Each frame contains a random number of cosmic ray and astrophysical X-ray events, with the numbers of events drawn from Poisson distributions corresponding to the anticipated mean event rates (one cosmic ray\cite{miller_jatis} and one X-ray event per frame, corresponding to a relatively low surface brightness source). X-rays are simulated with a uniform distribution of positions across the detector, applicable to observations of extended, low surface brightness X-ray sources, such as the intracluster medium, or ICM, of a galaxy cluster, or the cosmic X-ray background, and with a log-uniform distribution in energy between 0.3 and 10\,keV. An example of a simulated frame is shown in Figure~\ref{fig:sims:frame}.

While in real DEPFET and CCD detectors, the charge produced by the interaction of a single X-ray photon or charged particle can diffuse amongst neighboring pixels, producing single, double and quadruple pixel events, depending upon the pixel size, the \textsc{geant4} simulations do not incorporate this charge diffusion. For consistency we therefore assume that all X-rays are single-pixel events, although charge diffusion will be incorporated in future work.

\section{Classification of sub-frame regions of interest}
\label{sec:frame_class}
The first stage of the machine learning cosmic ray detection and filtering scheme is the identification of regions of interest within the frame images that are read out from the detector. Regions of interest are sub-frames, within the image, that are identified by a machine learning algorithm as likely containing (1) only astrophysical X-ray signals, (2) only signals induced by cosmic rays particles and their associated secondaries, or (3) a combination of astrophysical X-ray and cosmic ray events.

The baseline region size is defined as $64 \times 64$ pixels, although a hierarchy of region sizes can be applied in series to identify tracks and showers of secondary particles on different scales.

A sliding window is run over the frame image, extracting every possible $64\times 64$ region. The sub-frame images from the regions that contain at least one non-zero pixel are then classified using a convolutional neural network (CNN) algorithm\cite{spie_2020}. This algorithm employs two convolutional layers; kernels that are convolved with the image that represent the identifying features of the image that are learned. After the convolutional layers, the resulting layer is down-sampled by `maxpooling' (taking the maximum value in each $2\times2$ group of pixels). The layer applied after this down-sampling identifies  larger-scale features in the image. Finally, two fully-connected or dense layers are used to translate the output of the series of convolutional layers into the network's prediction. Following the common practice in machine learning algorithm development, a `dropout' layer is used to break some of the connections within network to mitigate over-fitting of the training data and to improve the accuracy of the algorithm when applied to new input data. The neural network algorithm was built using the \textsc{keras} high-level library on the \textsc{tensorflow} framework \cite{tensorflow}, and the architecture of the algorithms is shown in Appendix~\ref{app:networks}.

For each region, the algorithm returns a prediction consisting of three values that sum to unity. These values can be interpreted as the probability that the frame belongs to each of the classifications, \textit{i.e.} containing (1) only X-rays, (2) only cosmic rays, or (3) both. The CNN is trained upon 1,000,000 simulated sub-frames, as described above, for which the true classification is known.

\begin{figure}[h]
    \centering
    \small
    \includegraphics[width=16cm]{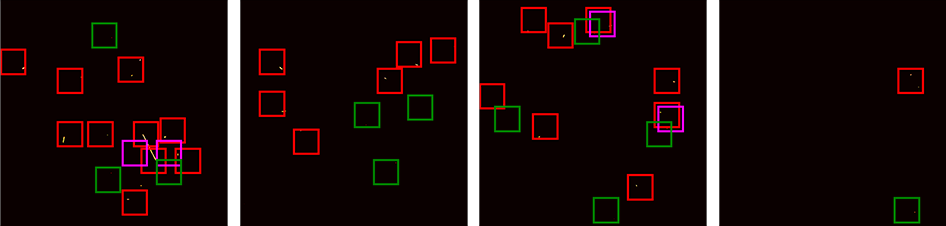}
    \caption{Regions of interest within simulated \textit{Athena} WFI frames, identified by the convolutional neural network image classification algorithm. The algorithm identifies regions containing only astrophysical X-rays (green boxes), only cosmic ray-induced events (red boxes), or both X-ray and cosmic ray events (magenta boxes).}
    \label{fig:regions}
\end{figure}

Where multiple possible regions of interest in each classification overlap, we employ non-maximal suppression. Only overlapping sub-frames with the highest prediction confidence in each category are kept. This produces the final list of regions of interest (Figure~\ref{fig:regions}). Those predicted to contain only X-rays can go forward for X-ray event reconstruction. Those predicted to contain only cosmic rays can be flagged and events contained in these regions can be rejected from the analysis. Finally, those predicted to contain both can be subject to further analysis to separate the X-ray and cosmic ray signals, described in \S\ref{sec:pixel_class}.

\subsection{Event energies}
\label{sec:energy}
Following the convention in machine learning image analysis, each image is normalised such that the maximum pixel value in each frame is unity. In X-ray detector frames, the pixel values represent the energy deposited (which, for X-ray photons, correspond to the energy of the photon). Normalising the images in this way prevents the algorithm `over-fitting', that is learning to identify specific frames based upon the specific energies of the specific events that happen to be within those frames in a way that does not generalise to all possible X-ray and cosmic ray events. 

The event energy, however, is an important factor in identifying X-ray and cosmic ray events. We therefore preserve the energy information contained within a frame by dividing the frame into \textit{channels}; a stack of images that contain only the pixels in specific energy ranges. The image in each channel is then normalised such that the maximum pixel value in each channel is unity. The convolutional filters act across the stack of $k$ channel images (\textit{i.e.} an $i\times j$ filter becomes an $i\times j\times k$ filter). The channels we employ in the prototype algorithm are 0.3-0.5\,keV, 0.5-1\,keV, 1-2\,keV, 2-5\,keV, 5-10\,keV and $>10$\,keV, although the channels can be tuned to optimize the performance of the algorithm.

\section{Pixel-by-pixel classification of detector frames}
\label{sec:pixel_class}
When regions of interest are identified as likely containing both astrophysical X-ray and cosmic ray signals, it is necessary to separate the signals within those regions such that the background signals can be filtered, but not at the expense of the astrophysical signals that are required for the scientific analysis.

These regions are input to a second machine learning algorithm that performs pixel-by-pixel classification and is based upon the \textsc{unet} architecture\cite{unet}. This algorithm first uses a series of convolutional filters to down-sample the image to a smaller number of predictions, similar to the CNN image classifier (referred to as the ‘encoder’ stage). A ‘decoder’ stage is then used to up-sample and map those predictions onto the original pixels of the image. In the prototype algorithm, the encoder stage contains four layers, with each layer operating a set of  $3\times 3$ convolutional filters on the image, followed by $2\times 2$ `max-pooling', to down-sample the image by selecting the largest pixel value within each $2\times 2$ region. The four layers contain 16, 32, 64 and 128 convolutional filters, respectively. The decoder stage operates the same architecture in reverse, concatenating the output of each decoded layer with the output of the encoder stage of corresponding size. Combining the step-by-step predictions with the original images from the encoder stage puts back the positional information that is lost with the downsampling in each layer. Finally, a $1\times 1$ convolutional layer is used to generate the prediction for each pixel. The number of layers and the number of filters within each layer can be tuned to optimize the performance of the algorithm.

The \textsc{unet} algorithm was trained on a further set of 1,000,000 simulated $64\times 64$ pixel sub-frames. In the training set (and hence in the \textsc{unet} output), we define a cosmic ray event as the entire rectangular region spanned by the entire shower of particles that reaches the detector from a single primary proton (\textit{i.e.} the cosmic ray truth value for all pixels in this region is 1.0). We find empirically that defining the cosmic ray event regions in this way, rather than only flagging the individual pixels illuminated by the particles, improves the performance of the algorithm and means that it is better able to associate the secondaries from a single event with one another. While such a definition means that valid X-ray events can be lost amongst the shower, we note that one would not typically believe that a valid event within the shower was a valid, astrophysical X-ray.

\begin{figure}[h]
    \centering
    \small
    \subfigure {
    \includegraphics[height=5.5cm]{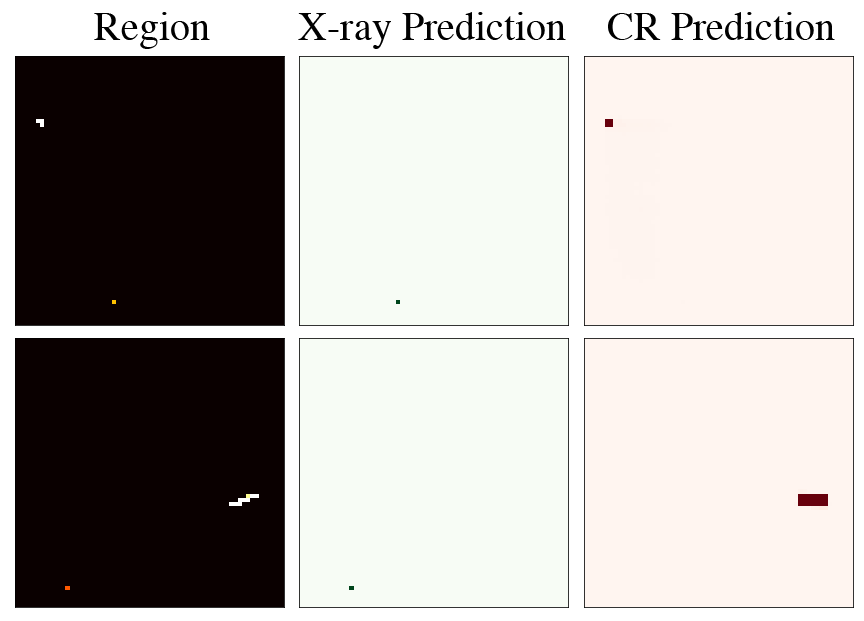}
    }
    \subfigure {
    \includegraphics[height=5.5cm]{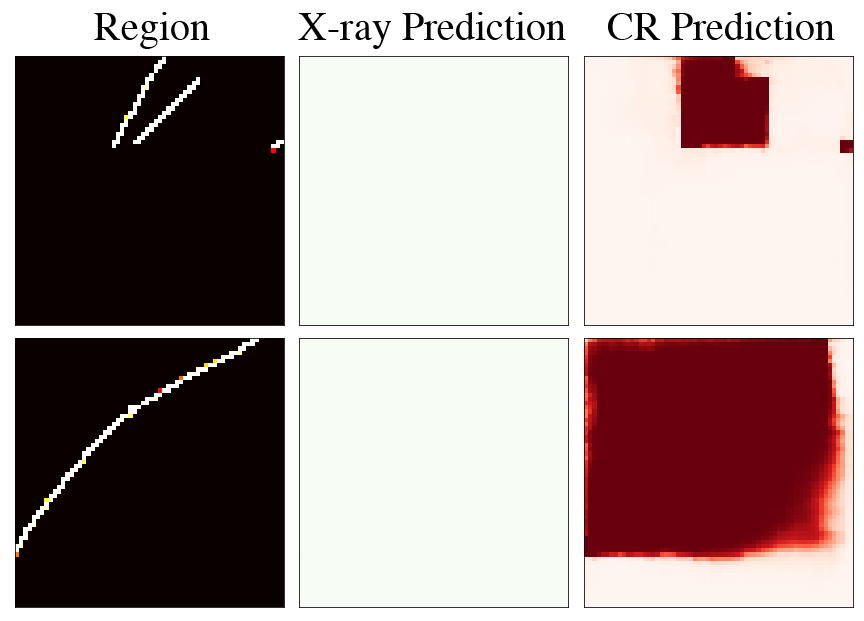}
    }
    \caption{Predictions from the \textsc{unet} pixel-by-pixel classification algorithm, run on regions of interest identified within the simulated \textit{Athena} WFI frames. For each pixel, predictions are computed, which can be interpreted as the probability that the pixel contains an X-ray or a cosmic ray-induced event, shown by the shading in the centre and right columns.}
    \label{fig:pixel}
\end{figure}

For each pixel, the \textsc{unet} algorithm returns three predictions (Figure~\ref{fig:pixel}), which can be interpreted as the probability that (1) the pixel is empty, (2) the pixel contains a signal from an astrophysical X-ray, and (3) the pixel contains a cosmic ray-induced signal.

X-ray events are then reconstructed from the frame image using the ASCA grading scheme to produce the event list. They are associated with the X-ray and cosmic ray probability values calculated by the Unet for the pixels in which they are detected (where an event spans multiple pixels, we associate the reconstructed event with the highest probability). The algorithm adds a column to the event list containing the probability that each event was induced by a cosmic ray and events exceeding a threshold cosmic ray probability can be filtered from the data set prior to commencing the scientific analysis.

\section{A hybrid AI background filtering algorithm}
\label{sec:hybrid}
The above stages are combined into a hybrid event classification algorithm, developed initially for the \textit{Athena WFI}, but which can be readily adapted for both existing and next-generation pixelated X-ray detectors. The input to the algorithm is a single frame image from one $512\times 512$ pixel quadrant of the detector, which is first split into a stack of frame images for the six energy channels described in \S\ref{sec:energy}. The first stage of the algorithm identifies regions of interest, using the sub-frame image classication CNN (\S\ref{sec:frame_class}). Events from regions identified as containing only cosmic rays are rejected, and regions identified as containing both cosmic ray and X-ray events are input to the pixel-by-pixel classification Unet algorithm (\S\ref{sec:pixel_class}), creating a pixel-wise map of cosmic ray probabilities. The ASCA-style grading algorithm is then used to identify and reconstruct events and each event is associated with the maximum cosmic ray probability computed by the algorithm over its $3\times 3$ pixel area, adding a `cosmic ray probability' column to the event list.

We test the performance of the hybrid algorithm on a set of 40,000 simulated frames, constructed using \textsc{geant4} events that were not part of the training sets. These frames contain a total of 126,000 events. We compare the performance of the algorithm to the filtering method employed on present-day X-ray detectors, using just the ASCA grading scheme, and rejecting background based upon the total event energy and event pattern or grade (i.e. number and shape of illuminated pixels). Figure~\ref{fig:spec} shows the reduction in the unrejected cosmic-ray induced background as a function of energy.

\begin{figure}[h]
    \centering
    \small
    \includegraphics[width=14cm]{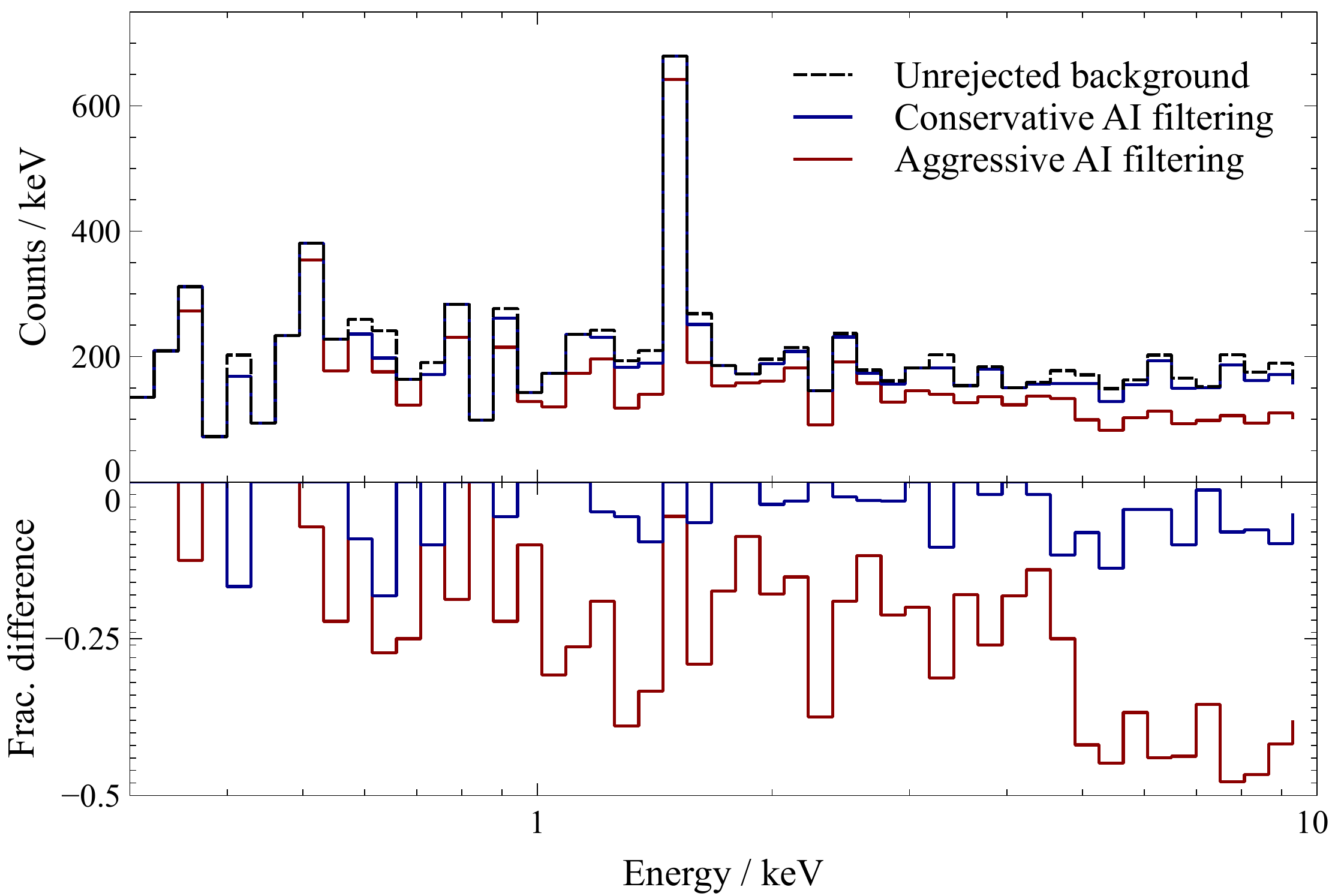}
    \caption{Simulation of the unrejected, cosmic ray-induced background spectrum in the \textit{Athena} WFI, as a function of detected energy. The unrejected background includes all events that have valid energies, in the range 0.3-10\,keV, and valid patterns (single or double pixel events) that would not be removed by present-day filtering criteria. The unrejected background is compared to the reduced  background achieved by the machine learning algorithm. The conservative mode removes all events in regions identified as containing only cosmic ray events, then employs pixel-by-pixel classification on regions identified as containing both cosmic ray and valid X-ray events. The aggressive mode rejects all events from $64\times 64$ pixel regions identified as containing both cosmic ray and X-ray events. The lower panel shows the fractional reduction as a function of energy when each of the methods is employed.}
    \label{fig:spec}
\end{figure}


We find that when employing the most aggressive mode of filtering, rejecting all events in regions identified to contain either cosmic rays or cosmic rays and X-rays, the machine learning algorithm provides a reduction in the unfiltered particle background by 30 per cent compared to the present-day method. This aggressive filtering, however, comes at the expense of reducing the effective area of the detector during each frame, removing the $64\times 64$ pixel regions containing the cosmic ray events. For reasonable X-ray event rates for low surface brightness sources, this translates to a loss of 5 per cent of valid X-ray events.

More conservative filtering, separating the pixels in regions identified to contain both cosmic rays and X-rays produces only a 6 per cent reduction in the background at the current time. However, the architecture of the pixel classification algorithm has not yet been optimised, and there is a feasible path to improving this performance.

96 per cent of the valid (\textit{i.e.} previously unrejected) background events that are correctly identified by the machine learning algorithm are found in frames with at least one other event, whether that is an invalid event, such as the particle track, or another valid event. We find that 3 per cent of the correctly-identified valid background events are found in frames with only other valid events, and are not accompanied by a particle track. We therefore conclude that machine learning algorithms are able to provide enhanced background filtering by learning the spatial correlations between the valid, unfiltered events caused by secondaries. These events can be associated with either the primary tracks or with other nearby valid events that result from the same primary. The high detector frame rate is essential, since this results in a small number of events per frame. This means that we know that for all but the brightest sources, there will only be of order one X-ray photon per frame, so that multiple valid events in close proximity can be identified as showers of secondaries resulting from a cosmic ray interaction.

\section{Conclusions}
\label{sec:conclusion}
A prototype machine learning algorithm has been developed to enhance the identification of background events in X-ray CCDs and similar imaging detectors that arise due to cosmic ray interactions with the spacecraft and detector. We employ a hybrid algorithm that first identifies regions of interest within a frame read out from the detector, using an image classification convolutional neural network. These regions are identified as containing either only astrophysical X-ray events, only cosmic ray-induced events, or both. Regions containing both X-ray and cosmic ray events are passed to a second stage algorithm, which classifies each pixel in the image, enabling the cosmic ray and X-ray events to be separated. We find that the prototype algorithm is able to reduce the instrumental background by up to 30 per cent compared with the present-day filtering method. Machine learning algorithms are able to consider every pixel in the frame while classifying events, and are able to learn the correlations between primary proton tracks and secondary events that enable otherwise valid events to be correctly identified as background.

\acknowledgments 
 
We thank Jonathan Keelan (Open University) and Rick Foster (MIT) for providing the \textsc{geant4} simulations of particle interactions with the \textit{Athena WFI} detector, in addition to routines and guidance for analyzing the \textsc{geant4} output. This work has been supported by the US \textit{Athena Wide Field Imager} Instrument Consortium under NASA grant NNX17AB07G, and the NASA \textit{Astrophysics Research and Analysis} (APRA) program under grant number  80NSSC22K0342.

\bibliography{ai_bkg} 
\bibliographystyle{spiebib} 

\clearpage
\appendix
\section{Machine learning algorithm architecture}
\label{app:networks}
\begin{figure}[h]
    \centering
    \small
    \includegraphics[height=14cm]{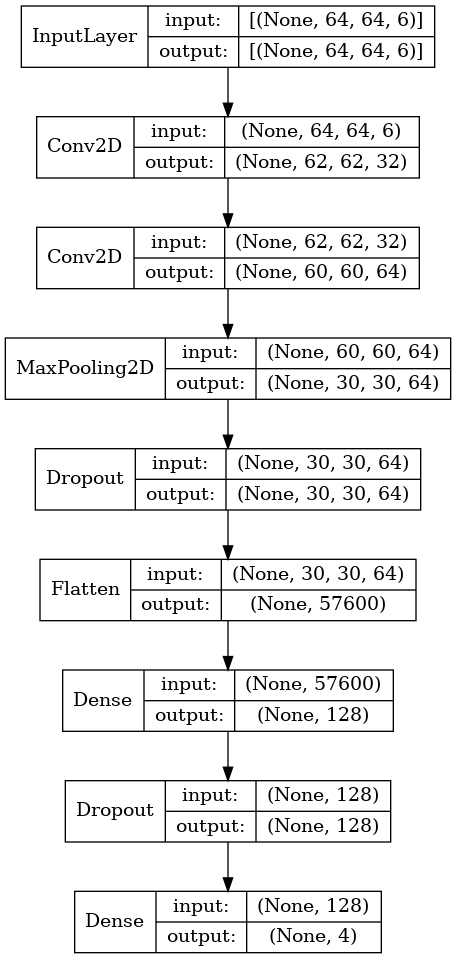}
    \caption{Architecture of the convolutional neural network algorithm to classify sub-frame images and identify regions of interest within the detector frames. The algorithm is constructed in \textsc{keras} to run on the \textsc{tensorflow} architecture. Given a $64\times 64$ pixel input image (separated into six energy channels), the algorithm returns three predictions, which can be interpreted as the probability that the frame contains (1) only astrophysical X-ray events, (2) only cosmic ray-induced events, and (3) both X-ray and cosmic ray events. A dimension of `None' indicates that the input size is flexible to process a set of images.}
    \label{fig:spec}
\end{figure}

\begin{figure}[h]
    \centering
    \small
    \includegraphics[height=21cm]{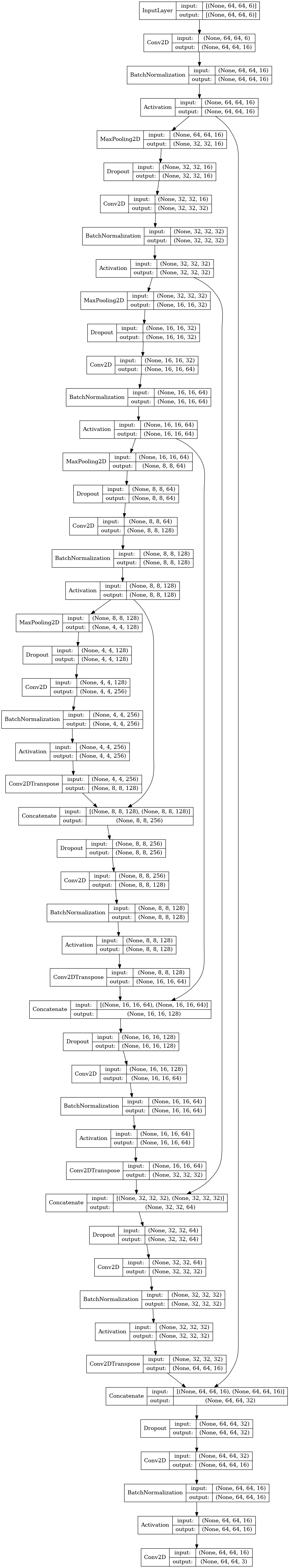}
    \caption{The Unet algorithm to perform pixel-by-pixel classification of regions of a frame identified as containing both X-ray and cosmic ray events. For each pixel, the algorithm returns three predictions, corresponding to the probability that the pixel (1) is empty, (2) contains signal from an X-ray event, and (3) contains signal from a cosmic ray event.}
    \label{fig:spec}
\end{figure}

\end{document}